\documentclass[preprint,12pt,authoryear]{elsarticle}

\usepackage{hyperref}
\usepackage{astrojournals}

\journal{New Astronomy}

\begin{document}

\begin{frontmatter}

  \title{On the role of magnetic fields in star formation}

  \author[label1]{C.~J.~Nixon\corref{cor1}} \ead{cjn@leicester.ac.uk}
  \author[label1,label2]{J.~E.~Pringle}
  \cortext[cor1]{Corresponding author}
  \address[label1]{Department of Physics and Astronomy, University of Leicester, Leicester, LE1 7RH, UK}
  \address[label2]{Institute of Astronomy, Madingley Road, Cambridge, CB3 0HA, UK}

\begin{abstract}
Magnetic fields are observed in star forming regions. However simulations of the late stages of star formation that do not include magnetic fields provide a good fit to the properties of young stars including the initial mass function (IMF) and the multiplicity. We argue here that the simulations that do include magnetic fields are unable to capture the correct physics, in particular the high value of the magnetic Prandtl number, and the low value of the magnetic diffusivity. The artificially high (numerical and uncontrolled) magnetic diffusivity leads to a large magnetic flux pervading the star forming region. We argue further that in reality the dynamics of high magnetic Prandtl number turbulence may lead to local regions of magnetic energy dissipation through reconnection, meaning that the regions of molecular clouds which are forming stars might be essentially free of magnetic fields. Thus the simulations that ignore magnetic fields on the scales on which the properties of stellar masses, stellar multiplicities and planet-forming discs are determined, may be closer to reality than those which include magnetic fields, but can only do so in an unrealistic parameter regime.
\end{abstract}


\end{frontmatter}

\section{Introduction} \label{sec:intro}
In two papers, \cite{Mestel:1965aa,Mestel:1965ab} argued for the importance of the role of magnetic fields in star formation. He pointed out that an average region of the interstellar medium (ISM) containing a stellar amount of mass cannot simply collapse to stellar densities, because it contains too much angular momentum. He argued that magnetic fields are likely to play a vital role in removing that angular momentum. At the same time, he pointed out that the average region of interstellar medium containing a stellar mass also contains too much magnetic flux for it to be able to collapse to stellar densities. Therefore the magnetic field has to find a balance between enabling the removal of angular momentum, and itself escaping from the collapsing material. He proposed that ambipolar diffusion might provide such a mechanism (see also \citealt{Mestel:1956aa}).

In contrast, \cite{Bate:2012aa} started with a self-gravitating, turbulent cloud core of mass $M = 500M_\odot$, density $n \approx 10^5\,{\rm cm}^{-3}$ and temperature $T = 10\,{\rm K}$, and followed the subsequent evolution. He was able to reproduce the observed initial mass function, and also the observed properties of binary and multiple stars, for stars less than around a solar mass. Similar results were obtained by \cite{Krumholz:2012aa} using a grid based code. Moreover, \cite{Bate:2018aa} has shown that his simulations also produce plentiful and massive discs around his protostars, of the kind required for planet formation \citep{Nixon:2018aa} and beginning to be seen around the youngest (Class 0 and I) protostars \citep[e.g.][]{Tobin:2015aa,Perez:2016aa}. None of the simulations by \cite{Bate:2012aa,Krumholz:2011aa,Krumholz:2012aa} included magnetic fields.

In the light of all this McKee \citep{McKee:2017aa} commented: ``{\it How is that possible when it is known that magnetic fields...have a major effect in extracting angular momentum from the accreting gas? In fact, in our current understanding, magnetic fields are so effective at extracting angular momentum that many simulations of the formation of protostellar disks fail to produce disks nearly as large as observed.}''

In fact, McKee's comments illustrate very well the problem with magnetic fields. If we do not put them into the simulations, then we can get results quite close to the observations. But if we include magnetic fields, then we do not. Application of Occam's Razor suggests a simple conclusion. But the question then is: how do we reconcile this with the observed presence of magnetic fields in and around regions of star formation \citep[see the review by][]{Crutcher:2012aa}? It is this apparent contradiction that we address in this paper.

\section{Do we need magnetic fields?}
The presence and influence of magnetic fields has been thought to play a major role in two aspects of the star formation process. First magnetic fields are able to transfer angular momentum efficiently and so are a potential solution of  Mestel's angular momentum problem. Second, magnetic fields are able to provide additional support to cloud material against gravitational collapse, and so can mediate, and in particular reduce, the rate at which star formation can proceed in dense interstellar material. We discuss each of these in turn.

\subsection{Is there an angular momentum problem?} \label{sec:angmom}
The picture of star formation envisaged by Mestel was that of the formation of a single star, such as the Sun, from the monolithic gravitational collapse of an amount of interstellar material. This concept was later developed in more detailed form, with single core, monolithic collapse calculations leading to the view of star formation summarized in the review by \cite{Shu:1987aa} (and also promulgated  in reviews by \citealt{Stahler:2005aa}, and by \citealt{McKee:2007aa}). It is clear that if one views the star formation process in terms of forming one star at a time from the interstellar medium which is of necessity rotating, then the need for the removal of angular momentum from the forming protostar becomes paramount.

The problem with this approach from the point of view of star formation is that it always leads to the formation of single stars. This is not a good result for typical solar mass stars of which only $50 \pm 10$ per cent are single \citep{Raghavan:2010aa}.

In view of this it is possible to make the case \citep{Pringle:1989aa,Pringle:1991aa,Clarke:1991aa,Reipurth:2001aa} that, contrary to the single core collapse picture, the formation of binary (and multiple) stars is in fact the way to understand the formation of all stars. The point is that in order to account for the occurrence of numbers of binary and multiple systems it is necessary that essentially all stars have to form in the presence of companions. If all stars form in groups, then many of these will be ejected as single stars (see the reviews by \citealt{Zinnecker:2001aa}; \citealt{Reipurth:2014aa}). And given that single stars are in a minority, it follows that most stars must form in groups. The observational case for the veracity of this conclusion is reviewed by \cite{Lada:2003aa}. This leads to the current model of chaotic star formation crystallized by \cite{Bate:2012aa}.

In this picture, it is to be expected that the angular momentum problem is to a large extent overcome by gravitational interactions alone \citep[e.g.][]{Larson:2010aa}, and this expectation is confirmed by the simulations. Thus it is clear that while magnetic fields may be present, they are not required to solve Mestel's fundamental angular momentum problem of removing angular momentum from the interstellar medium. 

Note, however, that the presence of magnetic fields is likely required at some level in the very late stages in order to help drive the final stages of disc evolution and the formation of jets, although \cite{Hartmann:2018aa} make the case that the importance of disc magnetic winds may have been overestimated. The early stages of disc evolution occur while the disc is self-gravitating \citep[e.g.][]{Nixon:2018aa} and around 90 percent of the stellar mass is accumulated in this way. However, the late stages, involving angular momentum from the last few per cent of the stellar mass, and the inner disc regions, from where the proto-stellar jets are driven, both involve discs that are ionized enough to support dynamo activity (MRI). However, the magnetic fields in these instances are unlikely to have been dragged in by accreting material \citep{Lubow:1994aa}. Local dynamo activity, acting on seed fields, is capable of generating the necessary viscosity through MRI, as well as generating larger scale, sufficiently ordered fields, that can drive dynamic outflows \citep{Tout:1996aa,Fendt:2018aa}.\footnote{An important distinction here is that in contrast to hydrodynamic turbulence, MHD turbulence can give rise to an inverse cascade whereby it is able to generate magnetic fields on lengthscales much larger than the driving lengthscale of the turbulence.}

We conclude that the problem of removing angular momentum from interstellar material in order to allow the formation of stars does not require a significant presence of large-scale magnetic fields. Indeed, it has been widely demonstrated \citep{Li:1996aa,Myers:2013aa,Myers:2014aa,Li:2014aa,Tomida:2015aa,Hennebelle:2016aa,Masson:2016aa,Kuffmeier:2017aa,Kuffmeier:2018aa,Kuffmeier:2018ab, Gray:2018aa} that introducing additional angular momentum transport (by introducing magnetic fields to the calculations) leads to the two major problems mentioned by McKee:\\

(i) it is difficult to reproduce the observed number of stars that are in binary and multiple systems, let alone the properties of the systems, and\\

(ii) it is difficult to produce the fraction of stars with massive enough discs to give rise to planet formation. \cite{Winn:2015aa} find that at least one half of solar-type single stars have planetary systems; and to form planets the disc masses need to be well above the minimum mass solar nebula of around $\sim 0.01M_\odot$ \citep{Nixon:2018aa}.

\subsection{Is there a star formation rate problem?}
The original perception of molecular clouds was that they are self-gravitating, isolated long-lived entities \citep[e.g.][]{Solomon:1987aa,Blitz:1991aa,Blitz:1993aa}. In that picture the observed supersonic turbulent support of the cloud was necessary in order to prevent the high star formation rate that would result from the gravitational contraction of the cloud on its free-fall or dynamical timescale. Moreover, it was thought that the turbulence needed to be strongly magnetic in order to cushion the shocks and so prevent rapid dissipation of the turbulence \citep{Arons:1975aa,Lizano:1989aa,Bertoldi:1992aa,Allen:2000aa}. However, it turned out that inclusion of magnetic fields has a minimal effect on the dissipation rate of the turbulence \citep{Ostriker:1999aa,Mac-Low:1998aa}. This idea that magnetic intervention is required in molecular clouds in order to slow the rate of star formation is indeed still prevalent \citep{Ballesteros-Paredes:2005aa,Vazquez-Semadeni:2005aa,Padoan:2011aa,Federrath:2013aa,Myers:2014aa,Padoan:2014aa,Federrath:2016aa}.

In recent times, this picture of molecular clouds has given way to a realisation that molecular clouds are much more transient entities.

First, \cite{Elmegreen:2000aa}, and others \citep[for example][]{Beichman:1986aa,Lee:1999aa,Jessop:2000aa,Ballesteros-Paredes:1999aa} have given cogent observational arguments that the star formation within a giant molecular cloud (GMC) occurs within one or two crossing times of its formation, that is within a few Myr. Similarly comparisons of the ages of young clusters and their association with molecular gas both in our Galaxy \citep{Leisawitz:1989aa} and in the Large Magellanic Cloud \citep{Fukui:1999aa} indicate that the dispersal of a cloud in which star formation has occurred takes a time-scale of only $5-10$\,Myr.\footnote{Incidentally, it follows from these observations that, contrary to what is often assumed \citep{Walch:2015aa,Padoan:2016aa,Kortgen:2016aa} since the vast majority of massive main-sequence lifetimes of stars that give rise to supernovae, ie $M \ge 8 M_\odot$, are $\ge 5-10$\,Myr \citep{Crowther:2012aa}, supernova explosions cannot provide an internal source of turbulent energy in GMCs. It has also been shown that supernova explosions cannot provide an external source of turbulent energy either \citep[see for example][]{Seifried:2018aa}.} Thus, molecular clouds are far more ephemeral than was previously postulated, and therefore the rate of star formation within them cannot be as high as previously envisaged.

Second, it has become apparent that GMCs as a whole are not self-gravitating \citep{Heyer:2009aa,Dobbs:2011aa}.\footnote{This implies that the discussion of the properties of such clouds in terms of ``free-fall times''  \cite[e.g.][]{Padoan:2014aa} not only has no meaning, but stems from the previous outdated physical picture \citep[see also][]{Kennicutt:2012aa}.} Numerical simulations of the evolution of the interstellar medium within disc galaxies show that the denser regions (the giant molecular clouds) are dynamic and transient structures \citep{Dobbs:2013aa,Dobbs:2015ab,Baba:2017aa}. They are not isolated objects and their evolution is highly complex. The larger clouds  (where most of the star formation takes place) form by cumulation of smaller clouds as well as directly from the denser regions of the ISM (convergent flows), and tend to disrupt because of galactic shear and feedback from star formation \citep[cf.][]{Meidt:2015aa,Dobbs:2018aa}. They are predominantly not self-gravitating, except for small regions within the clouds which give rise to star formation events and, hence, disruptive feedback. None of these simulations contains magnetic fields, but nevertheless, the overall star formation rates in such models are in line with those observed \citep{Dobbs:2011ab}. 

Thus, we conclude that the idea that magnetic fields are required to play a dominant, or even significant, role within molecular clouds in order to moderate the star formation rate is no longer tenable.

\section{Numerical simulations of magnetic fields in turbulent molecular clouds} \label{sec:incmag}
We have argued above that the presence of significant magnetic fields within the dense, star-forming interstellar gas is not required to explain the observed general properties of star formation.

It is, however, clear \citep[see the review by][]{Crutcher:2012aa} that magnetic fields are to be found in almost all regions of dense molecular gas in which star formation is occurring. The field strengths appear to be significant, in that the magnetic energy density is a substantial fraction of the energy associated with internal turbulent (or random) cloud motions, but are not dominant in that they are not strong enough to prevent global gravitational collapse of the molecular complex. Since the internal cloud motions are typically highly supersonic (with Mach numbers around $10-20$) this implies that the mean magnetic energy density strongly exceeds the thermal energy density. For typical cloud parameters, in order for the Jeans mass (given by a balance between thermal pressure and self-gravity) to be around a solar mass, it is therefore necessary for the cloud material that is actually forming stars to have shed much of its original magnetic flux \citep[cf.][]{Lubow:1996aa}.\footnote{It is worth noting that other authors, for example \cite{Federrath:2012aa}, also argue that the magnetic field plays at most a weak role in determining the final stellar masses.} The main question then is how this is achieved.

There are many simulations in the literature of the effects of driven, supersonic, but trans-Alf\'enic, turbulence on the gas density and magnetic field structures within model molecular clouds \citep[e.g.][see the reviews by \citealt{Ballesteros-Paredes:2005aa} and \citealt{Padoan:2014aa}]{Padoan:1999aa,Padoan:2011aa,Lemaster:2009aa}. In these simulations it is found that the turbulent motions create a range of densities, with the most dense regions in the tail of the distribution being subject to gravitational collapse (and presumed star formation). These dense regions still contain appreciable magnetic flux ($\beta_{\rm mag} = P_{\rm gas}/P_{\rm mag} \sim 0.4$, \citealt{Padoan:2011aa}). In none of these simulations was it possible to consider the formation of individual stars, let alone multiple stars or planet-forming discs. 

To remedy this, a step in the direction of extending the simulations of \cite{Bate:2012aa} and \cite{Krumholz:2012aa} to include the presence of magnetic fields has been reported in a series of papers \citep{Kuffmeier:2017aa,Kuffmeier:2018aa,Kuffmeier:2018ab}. In these simulations \citep[cf.][see also \citealt{Myers:2013aa,Gray:2018aa}]{Padoan:2016aa} the initial conditions consist of uniform density and uniformly magnetized molecular cloud material which is stirred by driven turbulence for some 10 Myr. At that time self-gravity is introduced and the cloud as a whole becomes self-gravitating. Thereafter the denser regions are subject to gravitational collapse, delineated by the occurrence of sink particles. The eventual stellar masses are around a solar mass. At this stage the minimum grid cell size is $126\,$au, which is too large to resolve binary or multiple star formation (median binary separation $20-40\,$au; Fig.~7 of \citealt{Duquennoy:1991aa} and Fig.~13 of \citealt{Raghavan:2010aa}), let alone the presence of protostellar discs. Regions of about $(40,000)^3\,{\rm au}^{3}$ around a few (six or nine, depending on the paper) of the sink particles are focussed on in the calculation, and the evolution of these regions are then followed for a further $\sim 10^4\,$yr at higher resolution, down to a minimum cell size of $\sim 2\,$au, although a region of $(14)^3\,{\rm au}^{3}$ is excised around the sink particle itself. In \cite{Kuffmeier:2017aa} disc formation is reported, and it is found that in these flows the outward transport of angular momentum is predominantly magnetic, rather then gravitational. Of the six sink particles studied in more detail in \cite{Kuffmeier:2018aa}, only two are found to have steady massive discs ($M_{\rm disc} \sim 0.01M_\odot$ and $R_{\rm disc} \sim 50-100\,$au), and of these one forms a companion with a separation of $\sim 1500\,$au. Two have no disc at all. The evolution of these discs is followed in more detail in \cite{Kuffmeier:2018ab}, where it is shown that all of the discs are strongly magnetic, and do not fragment.\footnote{For example the simple binary star formation mechanism discussed by \cite{Bonnell:1994aa} whereby a companion is formed by the interaction between gravitational instability in the protostellar disc and continuing accretion, cannot work if the disc is strongly magnetic.}

Thus, taken at face value, these simulations imply that the ubiquitous presence of magnetic fields in the molecular gas which is collapsing to form stars, seems to prevent  the desired outcome in terms of both the nature and properties of the resultant stars and of the properties of protostellar discs required for planet formation.

\subsection{Additional physics}
It may be, of course, that other physical effects can ameliorate the problem. We discuss two possibilities here. But at the same time it is worth discussing the extent to which  sets of numerical simulations, using current computer resources, are capable of representing physical reality. We do this in Section~\ref{sims}.

\subsubsection{Turbulent diffusivity}
One way of enhancing the diffusivity is to appeal to turbulent motions within the magnetised gas which might lead to an enhanced value of the effective diffusivity. This concept (turbulent diffusivity) is appealed to in various other branches of astrophysics, for example accretion disc theory \citep{Shakura:1973aa} and galaxy disc dynamo theory \citep{Ruzmaikin:1988aa,Shukurov:2004aa}. In both these examples, the source and properties of the turbulence are readily identified (the magneto-rotational instability in accretion discs \citep{Balbus:1991aa}, and the observed turbulent motions in the ISM for galaxy dynamos). Various authors \citep[for example][]{Fatuzzo:2002aa,Kim:2002aa,Zweibel:2002aa} discuss the possible enhancement of the effective diffusivity by adding turbulence in the context of large-scale star formation. In addition the same concept, under the nomenclature of ``reconnection-diffusion'', has been introduced by \citep{Lazarian:2005aa} and applied to the final stages of collapse to form a star by \cite{Leao:2013aa}. However, in the case of star formation the source and properties of the small-scale turbulence required to provide the enhancement in the effective diffusivity are neither readily identified nor discussed. Indeed it is questionable as to whether such a source exists. It is further questionable as to whether what is observed is actually ``turbulence'' in the usual fluid sense (see Section~\ref{ICturb}).

\subsubsection{Ambipolar diffusion}
Ambipolar diffusion (or ion-neutral drift) was discussed by \cite{Mestel:1965aa,Mestel:1965ab} as a mechanism whereby gas in the final phase of collapse to form a star might be able to shed itself of magnetic field. This is because at this stage the gas can be dense enough and cold enough to be predominantly neutral; see however the additional points raised by \cite{Norman:1985aa}. On the large scale in molecular clouds it is recognised that the effect is small. For example, \cite{Balsara:2001ab,Balsara:2001aa} find that in this context ambipolar drift does not play a significant role, and note further the findings of \cite{Mouschovias:1991aa} that ambipolar diffusion is mainly important in the last stages of collapse. Many recent authors \citep[for example][]{Li:1996aa,Chen:2014aa,Masson:2016aa,Wurster:2016aa,Auddy:2017aa,Gray:2018aa,Vaytet:2018aa} have come to similar conclusions. In addition \cite{Heitsch:2014aa} also conclude that in molecular clouds as a whole, neither ambipolar diffusion nor turbulent diffusion is likely to control the formation of cores or stars.

\section{How realistic are the turbulent MHD simulations?}
\label{sims}
We consider the answer to this question in two parts.  First we consider the extent to which the numerical simulations are able to simulate the relevant physical properties of the cloud material. We show, as recognised by those undertaking the simulations, that they are not. We then discuss the consequences of this disparity. Second, we consider the initial conditions assumed for the simulations compared to the current picture of molecular cloud formation.

\subsection{Physical properties of the cloud material}
The numerical simulations of MHD turbulence in molecular cloud material are, of necessity, restricted by what is numerically possible. The two parameters of immediate relevance are the Reynolds number ($Re$) and the magnetic Prandtl number ($P_{\rm M}$). 

As noted by \citet[][see also \citealt{Kritsuk:2009aa}]{Kritsuk:2011aa} the relevant Reynolds number is given by $Re \approx u L/\nu$, where $u$ is the r.m.s. velocity in the turbulence, $L$ is the relevant length scale (of order the energy injection scale) and $\nu$ is the fluid kinematic viscosity. These authors note that the largest values of $Re$ that can be reached are typically $\sim 10^4$ whereas realistic values for molecular clouds can be as high as $\sim 10^8$. Physically what this implies is that the viscosity inherent in the numerical codes is too high by several orders of magnitude. The main effect of this is that the smallest scales likely to be present in the turbulence are severely overestimated in the output of the simulations.

The magnetic Prandtl number is given by $P_{\rm M} = \nu/\eta$ where $\eta$ is the magnetic diffusivity. For typical molecular cloud material \cite{Kritsuk:2011aa} find that we may expect $P_{\rm M} \approx 2 \times 10^5 (x_i/10^{-7}) (n/1000 {\rm cm}^{-3})^{-1} \gg 1$. Here $x_i$ is the ionization fraction and $n$ the particle number density. In contrast, numerical simulations without explicit viscosity and explicit magnetic diffusivity, and which therefore rely on the grid scale to control both viscosity and diffusivity, generally and naturally have $P_{\rm M} \sim 1$. Since the numerical codes overestimate the viscosity by factors of order $\sim 10^4$, and underestimate the magnetic Prandtl number by of order $\sim 2 \times 10^5$, it follows that they overestimate the magnetic diffusivity by factors of order $\sim 2 \times 10^9$. Thus, to a first approximation, the simulations overestimate the rate at which cloud material can both divest itself of, and acquire, magnetic flux by {\it almost ten orders of magnitude}. It would be surprising if such a disparity did not have serious consequences.

\subsection{Nature of the driven turbulence}
\label{natturb}
In the numerical simulations the freeing of material from magnetic field occurs through driven turbulence coupled with a large magnetic diffusivity. Thus it is no surprise that those regions, in which gravity is just able to overcome magnetic fields and so enable collapse, still have near maximal field strength, viz. $\beta_{\rm mag} \sim 1$. However, it is well known that the properties of MHD turbulence differ substantially between the $P_{\rm M} \sim 1$ and the $P_{\rm M} \gg 1$ regimes (\citealt{Schekochihin:2002aa,Schekochihin:2002ab})\footnote{This distinction has been shown to have important consequences in accretion disc instability theory \citep{Potter:2017aa}}.

In hydrodynamic turbulence, turbulent energy is put into the flow at large scales. The energy is then transferred through a cascade of eddy sizes down to the smallest eddies whose size is controlled by the magnitude of the viscosity, $\nu$ \citep[e.g.][]{Tennekes:1972aa}. The smaller the viscosity, i.e. the higher the Reynolds number, the larger the range of eddy sizes, i.e. the smaller the scales at which kinetic energy is turned into heat. In (driven) MHD turbulence, with $P_{\rm M} \sim 1$, so that $\nu \sim \eta$ magnetic energy loss and kinetic energy loss are able to take place at the same small scales. This is what is occurring in the simulations. However, when $\eta \ll \nu$, so that $P_{\rm M} \gg 1$, this is no longer the case. As shown in \cite{Schekochihin:2002aa,Schekochihin:2002ab} in their model of a kinematic dynamo there is much more power in the magnetic field structure at small scales. In effect the magnetic field is stretched and folded into long thin structures, and it is the thinness of the structures that enables the magnetic energy dissipation to take place.

Thus in the high magnetic Prandtl number regime, this gives rise to the concept that the magnetic field structure is better imagined as a series of flux ropes. These ideas have been applied by \cite{Baggaley:2009aa} to incompressible MHD . In their model of a fluctuating dynamo the magnetic field is confined to thin flux ropes, advected by turbulence. Dissipation of magnetic field occurs predominantly through reconnection of flux ropes; but note that once reconnection occurs, magnetic energy is reconverted to kinetic as the field configuration rearranges itself. A similar, but cruder, model for similar processes occurring in supersonic magnetic turbulence in (therefore highly compressible) molecular clouds was developed by \cite{Lubow:1996aa}. They argued that reconnection processes in a 3D geometry lead inevitably to the formation of closed loops of field. This creates O-type neutral points which then enable the field to diffuse and dissipate. This leads throughout the cloud to a steady generation of dense material which has been freed from the direct influence of any permeating magnetic field. They concluded that such material would preferentially be the material within the cloud that partakes in star formation. Similar ideas were formulated by \cite{Shu:1987ab}, and have been revisited by \cite{Lazarian:2005aa}, \cite{Krasnopolsky:2005aa} and by \cite{Heitsch:2014aa}.

We note that it might seem reasonable to assume that overestimating the diffusivity would lead to underestimating the effect of the magnetic field. However, while this is the case in regions of high field strength, this is not the case in regions of low field strength. Regions of low field strength are, in the simulations,  overwhelmed with large flux from the high strength regions due to artificially high (numerical and uncontrolled) diffusivity. In simulations that have too high a diffusivity, star formation will only proceed when the magnetic field is just low enough, meaning that all star formation takes place with near-maximal field strengths \citep{Padoan:2011aa}. We argue that this cannot happen in reality as the real diffusivity is much lower than applied in the simulations. Thus cloud material which is highly magnetic, cannot free itself from fields and so cannot form stars \citep[cf.][]{Kortgen:2015aa}. Conversely, the material which is able to form stars is that material which is non-magnetic (either because it was already not threaded by field when the cloud formed (see below), or because it managed to shed field by reconnection in high $P_{\rm M}$ turbulence). Such non-magnetic material cannot occur in the simulations, because if it were present, the artificially high (numerical) diffusivity would feed large magnetic flux back into it from neighbouring regions.

\subsection{Initial conditions and nature of the turbulence}
\label{ICturb}
We have noted that simulations of star formation within magnetic clouds generally assume that all of the cloud material is initially uniformly threaded with magnetic field, and that it is then then subjected to driven turbulence. It seems unlikely that either of these assumptions is correct.

\subsubsection{Initial magnetic field distribution}
As we have noted above, molecular clouds appear to be ephemeral objects\footnote{In this respect molecular clouds are much more like atmospheric clouds than the original proposers of the nomenclature envisaged.} which readily form and disperse on timescales comparable to their kinematic crossing times \citep{Dobbs:2013aa}. Most of the material within them is not dominated by self-gravity. It is only small portions of the cloud that are compact enough to be subject to self-gravity, and so are able to collapse and form stars. The material from which the clouds form is expected to be denser than average ISM material \citep{Pringle:2001aa,Dobbs:2012aa} but since it is less dense, and more highly ionized, than cloud material it is to be expected that the magnetic diffusivity of the pre-cloud material is even smaller. Thus there is no reason to assume that the material from which clouds form is uniformly threaded with magnetic fields. Indeed it is more likely that such material would contain a large range of flux to mass ratios. For this reason it seems quite plausible that when gravitational collapse sets in, although some of the collapsing material is threaded by magnetic fields, some of it may not be. If that were the case, it would be expected that star formation would be more likely to occur from the material least threaded by magnetic flux.

\subsubsection{Cloud turbulence}
Although the velocity dispersions observed within molecular clouds are usually referred to as ``turbulence'' is it not at all clear that the motions represent well-developed turbulence in the standard fluid dynamical sense \citep[e.g.][]{Batchelor:1953aa,Tennekes:1972aa}. We have already commented that such motions cannot be driven by supernovae from either inside or outside the clouds.

Indeed in view of the current picture of molecular clouds in terms of ephemeral entities formed in regions of converging ISM material, often driven together in the context of spiral arms, it seems more likely that the supersonic velocity dispersions generally assumed to be ``turbulence'' are the result of energy released in the formation process. \cite{Bonnell:2006aa} demonstrate that if two clouds of ISM material, each of which has a {\it non-uniform} density structure, and each of which has {\it zero velocity dispersion}, are made to collide in a shock then the effect of the original clumpiness is to give rise to a velocity dispersion within the post-collision gas. In the astronomical literature such a velocity dispersion is invariably referred to as ``turbulence''. Within these clouds both the time-scale for the decay of these motions, and the time-scale for forming stars, are comparable to the clouds' dynamical lifetimes. In this model there is no need for any internal or external continuous driving mechanism for the ``turbulence''.

It is important to stress that it is the clumpiness of the pre-collision gas which gives rise to the post-collision velocity dispersion. That clumpiness, well observed within the ISM, is of course generated by instabilities and energy sources within the ISM, presumably including supernovae. The idea that  clumpiness needs to be generated post-collision from a pre-collision smooth flow \citep[e.g.][]{Heitsch:2008aa,Banerjee:2009aa,Micic:2013aa,Fogerty:2016aa,Fogerty:2017aa} is unnecessarily restrictive.

It evident that in all of these pictures the initial conditions in a cloud at the onset of gravitational collapse (and subsequent star formation) are unlikely to be close to those generated by driven homogeneous turbulence as found in the simulations.

\section{Discussion and Conclusions}
We have considered the apparent contradiction between the relative success of those models of the late, dynamical stages of star formation that do not include magnetic fields, and the observed presence of magnetic fields within molecular clouds and cloud cores where stars form. 

We have argued that the earlier concept of star formation in terms of single star collapse, which led to the notion of an angular momentum problem (and therefore to the need for, and importance of, magnetic fields), has been replaced by the more recent concept of chaotic star-formation \cite[e.g.][]{Bate:2012aa}, where stars form predominantly in groups. In this picture gravitational interactions provide a solution to the angular momentum problem, except for the late-stage evolution of the inner regions of the protostellar discs where MHD turbulence is likely involved. \cite{Hartmann:2018aa} and \cite{Simon:2017aa} make the case that while there is little evidence for magnetic activity in the outer regions of protostellar discs, there is evidence of magnetic activity (magnetic winds, jets) in the inner regions of those discs, where the temperatures are high enough for an MRI-driven dynamo to be present. Because of velocity considerations (outflow velocities are comparable to escape velocities from the central object), it has long been argued that the major components of outflows are driven from close to the inner disc radii \citep{Konigl:1986aa,Pringle:1993aa,Livio:1997aa,Price:2003aa}. In addition the strongest protostellar outflows are found to occur among the youngest (strongly accreting, and often heavily embedded) objects \citep[e.g.][]{Bally:2016aa}. Magnetic winds and jets do require the presence of a global field, but there is no need for this to have been advected by the disc -- indeed that in itself is unlikely \citep{Lubow:1994aa}. In such strongly ionized disc regions (such as the inner regions, and in hot strongly accreting discs) it is possible for the MRI-dynamo itself to create a sufficiently large global field for jet-launching \citep{Tout:1996aa}.

We have also noted that the idea that magnetic fields play an important role on a larger scale, preventing the gravitational collapse of molecular clouds and slowing down the rate of star formation within them comes from the earlier concept that molecular clouds are self-gravitating isolated entities. The more modern view is that this is not the case. Thus magnetic fields are no longer needed to play a significant role in slowing down the star-formation process.

In this context, we have considered the ability of current numerical simulations to emulate the early stages of star formation from molecular material. We have noted problems with two aspects of this work. First, in many simulations the material is assumed to be uniformly threaded with magnetic field and then subjected to prolonged driven turbulence. We have argued that this may not be a good representation for the initial stages of gravitational collapse in a star-forming cloud. Second, and more seriously, we have noted that the physical conditions of the MHD being simulated (especially with regard to the magnetic Prandtl number and the magnetic diffusivity) differ between the simulations and physical reality by many orders of magnitude. In particular, the magnetic diffusivity, which provides the timescale on which cloud material is able to lose, and to acquire, magnetic flux, is overestimated by almost ten orders of magnitude. Since the region of parameter space (in, for example, the magnetic Prandtl number -- diffusivity plane) that represents physical reality is so far removed from what is amenable to numerical simulation, it is reasonable to question the usefulness of proceeding along these lines. In any case it is clear that those papers which present such numerical simulations do need to include some justification and discussion of the extent to which such simulations can be expected to represent physical reality.

In view of all this, we have advanced the hypothesis that there is a much larger scale of flux to mass ratios present in the relevant molecular material than can, at present, be simulated numerically. If so, we suggest that it would be predominantly the material in the cloud that is relatively free of magnetic field that partakes in the formation of stars \citep[cf.][]{Lubow:1996aa}. Thus simulations that ignore magnetic fields on the scales on which the properties of stellar masses, stellar multiplicities and planet-forming discs are determined, may be closer to reality.

\section*{Acknowledgments}
We thank Clare Dobbs, Charles Gammie, Lee Hartmann, Michael K{\"u}ffmeier and Chris McKee for useful correspondence. We thank Bob Carswell and Andrew King for support and encouragement. CJN is supported by the Science and Technology Facilities Council (grant number ST/M005917/1).

\bibliographystyle{elsarticle-harv}
\bibliography{nixon}

\end{document}